\documentclass[twocolumn,floatfix, showpacs]{revtex4}

\usepackage[final]{epsfig}
\usepackage{amsmath}
\begin{document}
 
\title{The Rapidity Dependence of Jet Quenching }
 
\author{Thorsten Renk}
\email{thorsten.i.renk@jyu.fi}
\affiliation{Department of Physics, P.O. Box 35, FI-40014 University of Jyv\"askyl\"a, Finland}
\affiliation{Helsinki Institute of Physics, P.O. Box 64, FI-00014 University of Helsinki, Finland}

\pacs{25.75.-q,25.75.Gz}

\begin{abstract}
The suppression of high transverse momentum ($P_T$) jets and hadrons in ultrarelativistic heavy-ion collisions with respect to a p-p baseline in terms of the nuclear suppression factor $R_{AA}$ is one of the key observables to gauge the density of a hot and dense QCD medium. However, the suppression measured by $R_{AA}$ is not a straightforward measure of the medium properties, the value of the observable also depends on the ratio of quark to gluon jets and on the slope of the hard parton spectrum, which explains why $R_{AA}$ is found to be fairly similar at RHIC and LHC despite the very different dynamics. Measuring high $P_T$ jets and hadrons at forward rapidity offers the same possibility of varying medium density, parton mixture and spectral slope without the need to compare across different $\sqrt{s}$ and experiments. In this work, the well-tested jet quenching Monte-Carlo (MC) framework YaJEM is utilized to compute the rapidity dependence of $R_{AA}$ for three test cases. 
\end{abstract}
 
\maketitle

\section{Introduction}

The idea to utilize high $P_T$ processes as a tomographic probe to study the density evolution of the Quantum Chromodynamics (QCD) matter created in ultrarelativistic heavy ion collisions \cite{radiative1,radiative2,radiative3,radiative4,radiative5,radiative6} has now led to a rich experimental program at both the Brookhaven Relativistic Heavy Ion Collider (RHIC) and at the CERN Large Hadron Collider (LHC). The classic observable in this context is the nuclear suppression factor $R_{AA}$ which is defined as the yield of high $P_T$ probes at rapidity $y$ from an A-A collision normalized to the yield in p-p collisions at the same energy corrected for the number of binary collisions, 

\begin{equation}
\label{E-RAA}
R_{AA}(P_T,y) = \frac{dN^h_{AA}/dP_Tdy }{T_{AA}({\bf b}) d\sigma^{pp}/dP_Tdy}.
\end{equation}

Initially measured at RHIC for high $P_T$ charged hadrons \cite{PHENIX-classic}, $R_{AA}$ is now measured with greater precision and more differentially for hadrons at both RHIC \cite{PHENIX-new} and LHC \cite{ALICE-RAA,CMS-RAA}. For fully reconstructed jets, preliminary results exist for $R_{AA}$ \cite{ALICE-jet-RAA} and measurements of a similar quantity $R_{CP}$ (which is the ratio of central over peripheral p-p like heavy ion collisions) have been published\cite{ATLAS-jet-RCP,ALICE-jet-RCP}.

However, $R_{AA}$ is not a pure tomographic probe in the sense that it would measure the density of the medium directly, as for instance evident from the initial surprise that the charged hadron $R_{AA}$ measured at LHC \cite{ALICE-RAA} was very similar to the RHIC value \cite{PHENIX-new}, despite the substantially higher bulk multiplicity production seen at the LHC indicating a larger medium density. It was soon widely realized that the observed suppression depends not only on the opaqueness of the medium to high $P_T$ probes, but also on the relative mixture of quark to gluon jets (with the gluons coupling by a factor $C_F = 9/4$ more strongly to the medium due to their different color charge) and the slope of the hard parton spectrum prior to the interaction with the medium (with a harder spectrum, more energy needs to be lost from the leading parton to achieve the same amount of suppression). For a pedagogical introduction of these effects, see e.g. \cite{Kinlimit}.

Measuring high $P_T$ probes at forward rapidity are an interesting way of probing the role of these effects within the same experiment, as spectral slope, parton type mixture and medium density do not only vary with $\sqrt{s}$ but equally well as a function of the rapidity $y$. Thus, measurements of $R_{AA}(y)$ (such as currently done by the ATLAS collaboration \cite{ATLAS-RAA-y}) can complement information on the $\sqrt{s}$ dependence of jet quenching established across RHIC and LHC experiments. Model predictions for RHIC kinematics exist e.g. in the Arnold-Moore-Yaffe (AMY) framework \cite{AMY-forward}, however at present the rapidity coverage of the high-$P_T$ capable detectors is limited. 

The aim of this work is to present case studies for $R_{AA}(y)$ at LHC kinematics for both hadron and jet suppression as a function of rapidity utilizing the well-constrained in-medium shower evolution code YaJEM \cite{YaJEM1,YaJEM2}. However, as will be argued later, qualitatively the results do not strongly depend on specifics of the parton-medium interaction model used but are rather driven by properties of the primary parton spectrum computed in perturbative QCD (pQCD).

\section{Basic considerations}

Observing the yield of high $P_T$ probes at fixed transverse momentum at increasingly forward rapidities corresponds to a fast approach to the kinematic limit. Since for a produced parton $E = p_T \cosh y$ (neglecting parton mass) and $E_{max} = \sqrt{s}/2$, the kinematic limit for parton production in 2.76 ATeV Pb-Pb collisions at $y=3$ corresponds to a mere 137 GeV. As this limit is approached, two important changes happen with respect to the situation at midrapidity at the same $P_T$ : 1) the slope of the parton spectrum steepens, if a local power law fit of the spectrum of the form $1/p_T^n$ is made, the power $n$ grows and 2) kinematics forces to probe the high $x$ region in the initial parton distribution functions (PDFs)\cite{CTEQ1,CTEQ2} which corresponds to the valence quark distribution, i.e. a mixture of quark and gluon jets at midrapidity changes to a pure quark jet distribution at forward rapidity.

With regard to the suppression measured by $R_{AA}$, these two effects act in oppsite directions. In order to see this qualitatively, consider a simple toy model in which the effect of the medium is approximated by a mean energy loss $\langle \Delta E \rangle$. In this case, the energy loss shifts the parton spectrum as partons travserse the medium. This can be described by the replacement $p_T \rightarrow p_T + \langle \Delta E \rangle$ in the expression for the parton spectrum after partons exit the medium. $R_{AA}(p_T)$ can then be approximated by the ratio of the parton spectra before and after energy loss as

\begin{equation}
\label{E-RAAApprox1}
R_{AA}(p_T) \approx \left(\frac{p_T}{p_T + \langle \Delta E \rangle}\right)^n = \left(1 - \frac{\langle \Delta E \rangle}{p_T + \langle \Delta E \rangle}\right)^n.
\end{equation}

If the spectrum gets steeper, $n$ increases and, for fixed $\langle \Delta E \rangle$, $R_{AA}$ \emph{decreases}. However, since due to the different color factor $\Delta E_{gluon} \approx 9/4 \, \Delta E_{quark}$, decreasing the fraction of gluon jets decreases $\langle \Delta E \rangle$, which in turn \emph{increases} $R_{AA}$. The net result of moving to forward rapidity will hence depend on how well the cancellation between these different effects takes place in a realistic framework.

On top of these purely kinematical effects is a (small) genuine variation in the density of QCD matter with rapidity. While for $y<2$ the rapidity dependence of multiplicity is measured to be rather flat \cite{CMS-dNdy}, towards $y=3$ a 10-20\% reduction is apparent \cite{ALICE-dNdy}, leading to a corresponding reduction in $\langle \Delta E \rangle$. Note however that this is parametrically a small effect when compared to the changes in the primary parton production spectrum: In 2.76 ATeV collisions at 100 GeV, about 70\% of all partons are gluons at $y=0$ whereas the fraction drops to 15\% at $y=3$, leading to a 40\% reduction of $\Delta E$ in the simple model above. 

\section{The model}

In reality, the pQCD parton spectrum is not a simple power law, parton mixture varies as a function of $p_T$ as well as with rapidity, the medium effect is not a simple and constant shift of parton energy but a probabilistic modification of the fragmentation function dependent on the actual path of the parton shower through the medium and $R_{AA}$ is computed for an observed object (either a hadron or a jet) which has only a probabilistic relation to the momentum of the parent parton  whose shower it originated from.

In order to take these complications into account, a leading order pQCD computation is used to obtain the parton spectrum as a function of $p_T$ and $y$, followed by a medium modified final state shower computed with the Monte-Carlo (MC) code YaJEM \cite{YaJEM1,YaJEM2} embdedded into a fluid dynamical simulation of the bulk medium. The computation for the mid-rapidity case for both charged hadron $R_{AA}$ and jet $R_{AA}$ using the anti-$k_T$ algorithm \cite{FastJet} with $R=0.3$ as jet definition is described in detail in \cite{YaJEM-RAA} and the reader interested in the details is referred to this work.

The extension of the LO pQCD parton production computation to forward rapidity is straightforward. A somewhat greater challenge is posed by extrapolating the boost-invariant fluid dynamical medium description \cite{RAA-LHC} used in \cite{YaJEM-RAA} consistently to larger rapidities. For the following computations, guided by the evolution of the charged particle pseudorapidity distribution \cite{ALICE-dNdy} the density beyond $y=2$ is smoothly reduced by 20\% towards $y=3$. Based on comparing with the case of not changing the medium at all, a conservative 10\% systematic error is assigned to account for effects missed by this procedure, for instance possible changes in the geometry of matter distribution. As indicated above, in comparison to the kinematic effects, the role of the rapidity dependence of the medium density is comparatively small.

\section{Results}

\begin{figure}[htb]
\epsfig{file=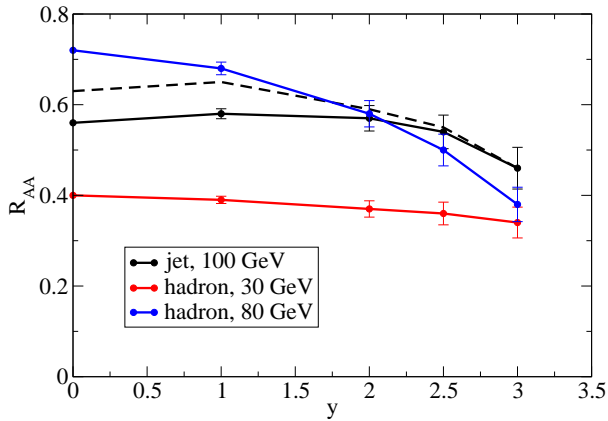, width=8cm}
\caption{\label{F-1} Nuclear suppression factor $R_{AA}$ for 30 and 80 GeV charged hadrons and 100 GeV $R=0.3$ anti-$k_T$ jets as a function of rapidity in 2.76 ATeV 0-10\% central Pb-Pb collisions computed with YaJEM. The dashed line indicates the result assuming all showers evolve and interact with the medium as if originating from a quark.}
\end{figure}

In the following, the rapidity dependence of $R_{AA}$ in the range of 0 to 3 units is studied for three different cases: charged hadrons at 30 GeV (far from the kinematic limit) and 80 GeV (approaching the kinematic limit at forward rapidity) and 100 GeV jets clustered with the anti-$k_T$ algorithm at a radius of $R=0.3$ (also close to the kinematic limit at forward rapidity). Note that at midrapidity, the results are in agreement with data \cite{YaJEM-RAA}. The rapidity dependence of these scenarios, which is the main result of this work, is shown in Fig.~\ref{F-1}. For illustration of the role of the parton mixture change, a line has been added to show the result if only quark fragmentation would contribute to 100 GeV jets everywhere.

As can be seen from the figure, the rapidity dependence is to first order remarkably flat, with a consistent trend of a drop of $R_{AA}$ with increasing rapidity, seen strongest in the case of 80 GeV hadrons. This indicates that ultimately the kinematic effect of spectral steepness wins out, especially close to the kinematic limit. As evidenced by the test case with quark fragmentation only, the changing parton mixture partially counteracts this trend as expected. The observed flatness of jet $R_{AA}$ out to 2 units of rapidity is in good agreement with the ATLAS measurement \cite{ATLAS-RAA-y} (which however utilizes a somewhat different jet definition, affecting mainly the absolute normalization of $R_{AA}$ rather than $y$ dependence).

\section{Discussion}

In order to gain more insight into the physics determining these results, it is useful to discuss the situation in terms of biases as outlined in \cite{Bias}. In this view, finding a hadron or a jet at given $P_T$ and $y$ represents a condition, and we can study how kinematic probability distribution (such as the conditional probability of a gluon jet, or the conditional probability for having a certain parton momentum) depend on the condition and respond to the bias introduced by the medium.

\begin{figure}[htb]
\epsfig{file=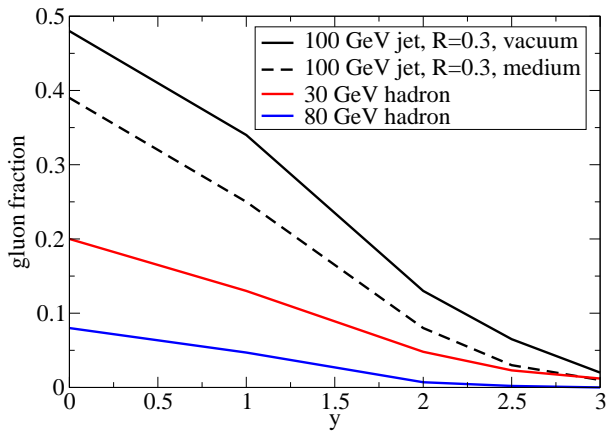, width=8cm}
\caption{\label{F-2}The fraction of gluons contributing to 100 GeV anti-$k_T$ jets with $R=0.3$ both in vacuum and in the medium created in 2.76 ATeV 0-10\% central Pb-Pb collisions and of hadrons at 30 and 80 GeV in vacuum (medium modifications are small in the hadronic cases).}
\end{figure}

In Fig.~\ref{F-2}, the fraction of gluon jets contributing to the respective observable is shown. First of all, it is evident that the gluon fraction in the primary parton spectrum is not a good proxy for the gluon fraction contributing to the observend distribution of hadrons and jets, i.e. the parton type bias induced by the observable definition is strong. While gluon jets contribute at midrapidity quite significantly to the jet yield, their contribution to the hadron yield is much reduced due to the fact that gluons have an on average softer fragmentation pattern than quarks. Even at comparatively low $P_T$ of 30 GeV, about 80\% of hadrons come from quark jets. Thus, while the change of the relative parton contributions with rapidity is an important effect for jets, it clearly is a subleading effect for hadrons. This partially explains why jet and hadron $R_{AA}$ for 80 GeV become more similar if all partons are assumed to be quarks in the jet case.

\begin{figure}[htb]
\epsfig{file=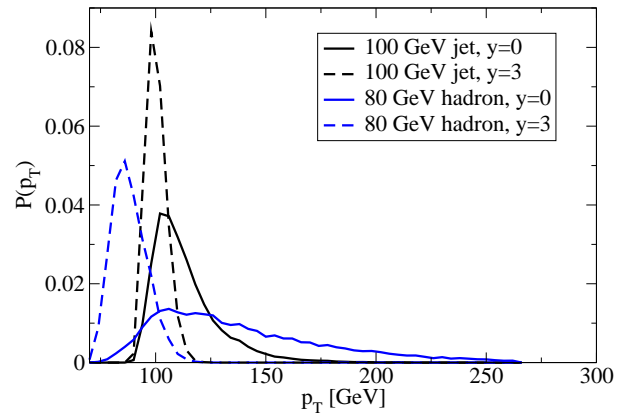, width=8cm}
\caption{\label{F-3}Parton kinematics contributing to observed hadrons at 80 GeV and  to 100 GeV anti-$k_T$ jets with $R=0.3$ at 2.76 ATeV both at $y=0$ and $y=3$.}
\end{figure}

However, if we attribute the downward trend at high $y$ to the steepening of the parton spectrum close to the kinematic limit, it is at first glance still surprising that 80 GeV hadrons are much more strongly affected than 100 GeV jets. This can nevertheless be understood by considering the underlying parton kinematics as shown in Fig.~\ref{F-3}.

Jets at 100 GeV are fairly collimated objects and the clustering, even with a comparatively small radius $R=0.3$, captures a large fraction of the original parton energy, leading to a well-defined relation between jet and parton kinematics --- jets at 100 GeV rarely originate from partons above 150 GeV. This is not the case for hadrons at 80 GeV, where at $y=0$ significant contributions still come from partons above 200 GeV. Going to $y=3$ restricts the phase space, as discussed before, to partonic $p_T < 137$ GeV. While this restriction is felt by the jets, as evident by the shift of the conditional probability, it is a much stronger effect for hadrons, where a significantly part of the phase space contributing at $y=0$ is no longer accessible at $y=3$, resulting in a drastic change in the shape of the probability distribution. This difference in parton kinematics explains why hadrons turn out to be actually more sensitive to the kinematic limit than jets, even with the jet energy being nominally 20 GeV closer to the limit.

\section{Conclusions}

Measuring the suppression of high $P_T$ objects at forward rapidities is a tool to study the response of $R_{AA}$ to changes in the relative mixture of quarks and gluons and to the slope of the partonic spectra, and to a lesser degree to changes in the opaqueness of the medium. A similar set of factors is explored by studying high $P_T$ observables as a function of $\sqrt{s}$. Going to forward rapidities complements such measurements rather than uncovering novel physics. To some degree, the contributing factors in terms of kinematics probed and parton mixture explored can be experimentally controlled by designing the observable --- different jet definitions can make the physics more parton-like (in terms of clustering almost all energy of the original parton back into the jet) or more hadron-like. Compared with the rapidity dependent kinematical effects, typical medium-induced biases are small (see Fig.~\ref{F-2} or \cite{Bias}), hence pQCD for p-p collisions is a reliable estimator for the kinematical conditions under which jet quenching happens.

The expected effect magnitude is unfortunately small unless one reaches close to the kinematic limit, which means that high statistics will be required on the experimental side to conclusively verify whether the downward trend described in this work is found in nature as well. It is however a truism that any study of the response of jet quenching observables to a more steeply falling parton spectrum, be it by going to low $\sqrt{s}$ or by approaching the kinematic limit, will be statistics-hungry by definition. It should be noted however that the fact that a flat $R_{AA}$ in rapidity over a wide range is expected is far from trivial, as the prior distribution of partons changes significantly in this range, i.e. an unchanged $R_{AA}$ can not be taken as an indication of the same physics, but must be seen as a (partial) cancellation of opposing effects.

Potentially more differential information could be obtained by studying hard jet or hadron back-to-back correlation as a function of the rapidity of both trigger and associated object. An assessment of the physics potential of such a study will be left for future work.

\begin{acknowledgments}
  This work is supported by the Academy researcher program of the
Academy of Finland, Project No. 130472. 
 
\end{acknowledgments}

\end{document}